\RequirePackage{fix-cm}
\documentclass[twocolumn,epjc3]{svjour3}  
\smartqed  
\RequirePackage{graphicx}
\journalname{Eur. Phys. J. C}
\usepackage[lofdepth,lotdepth]{subfig}
\usepackage{amsmath}
\usepackage{upgreek}

\def\Xmax{$X_\mathrm{max}$}
\def\XmaxSpace{$X_\mathrm{max}$~}
\def\XmaxMath{X_\mathrm{max}}
\def\Rhomu{$\rho_{\mu}^{600}$~}
\def\Srd{$S_{\textrm{RD}}^{\rho_{\theta}}$~}
\def\sqrtSrd{$\sqrt{S_{\textrm{RD}}^{\rho_{\theta}}}$~}

\def\RhomuSrd{$\rho_{\mu}^{600} / \sqrt{\textrm{S}_{\textrm{RD}}^{\rho_{\theta}}}$~}
\def\obslev{1552\,m a.s.l. }

\usepackage[disable]{todonotes} 
\usepackage{color}

\begin{document}

\graphicspath{{figs/}}

\newcommand{\degree}[1]{#1$^{\circ}$}

\title{Enhancing the cosmic-ray mass sensitivity of air-shower arrays by combining radio and muon detectors}


\author{Ewa M. Holt\thanksref{e1,addr1,addr2}
        \and
        Frank G. Schr\"oder\thanksref{e2,addr1,addr3} 
        \and
        Andreas Haungs\thanksref{addr1}
}

\thankstext{e1}{e-mail: ewa.holt@kit.edu}
\thankstext{e2}{e-mail: fgs@udel.edu}


\institute{Institut f\"ur Kernphysik, Karlsruhe Institute of Technology(KIT), Karlsruhe, Germany \label{addr1}
           \and
           Instituto de Tecnolog\'ias en Detecci\'on y Astropart\'iculas (CNEA, CONICET, UNSAM), Buenos Aires, Argentina \label{addr2}
           \and
           Bartol Research Institute, Department of Physics and Astronomy, University of Delaware, Newark, DE, USA \label{addr3}
}

\date{Received: date / Accepted: date}

\maketitle

\begin{abstract}
The muonic and electromagnetic components of air showers induced by cosmic-ray particles are sensitive to their mass.
The sizes of the components can be measured with particle detectors on ground, and the electromagnetic component in addition indirectly via its radio emission in the atmosphere. 
The electromagnetic particles do not reach the ground for very inclined showers.
On the contrary, the atmosphere is transparent for the radio emission and its footprint on ground increases with the zenith angle.
Therefore, the radio technique offers a reliable detection over the full range of zenith angles, and in particular for inclined showers. 
In this work, the mass sensitivity of a combination of the radio emission with the muons is investigated in a case study for the site of the Pierre Auger Observatory using CORSIKA Monte Carlo simulations of showers in the EeV energy range.
It is shown, that the radio-muon combination features superior mass separation power in particular for inclined showers, when compared to established mass observables such as a combination of muons and electrons or the shower maximum \Xmax.
Accurate measurements of the energy-dependent mass composition of ultra-high energy cosmic rays are essential to understand their still unknown origin.
Thus, the combination of muon and radio detectors can enhance the scientific performance of future air-shower arrays and offers a promising upgrade option for existing arrays.



\end{abstract}

\section{Introduction}
\label{intro}

More than 100 years after the discovery of the cosmic rays, the origin of the ultra-high energy cosmic rays is still under investigations.
Recently, the Pierre Auger Collaboration found evidence for an extragalactic origin of cosmic rays above $8 \cdot 10^{18}$\,eV by measuring an anisotropic distribution of the arrival directions \cite{Aab:2017tyv}.
However, the sources accelerating these cosmic rays up to the highest energies are still unknown. 
To answer open questions about their sources and propagation mechanisms through studies of the arrival direction and the energy spectrum, accurate measurements of the mass composition are essential.
Cosmic rays above $10^{14}$\,eV are measured indirectly via extensive air showers in the Earth's atmosphere.
Thereby, the primary cosmic rays induce a cascade of secondary particles, of which mainly muons, electrons and photons arrive on ground and can be measured with particle detectors.
In addition, mainly the electrons induce fluorescence light, Cherenkov light, and radio emission in the atmosphere, which can be measured on ground \cite{Kampert:2012,Schroder:2016hrv,Huege:2016veh}. 
All three methods have comparable intrinsic sensitivity to the electromagnetic shower component, but optical detectors are restricted to clear and dark nights while radio antennas can be operated at almost any weather conditions. 
In the last few years radio arrays have achieved a measurement accuracy competitive to the optical techniques. 
This stimulated this research on possible synergies between radio and particle detectors, i.e., the two techniques allowing for continuous operation.
\par 
The development of the muonic and electromagnetic components in a cosmic-ray air shower depends on the mass of the primary particle. 
Since air showers induced by heavier particles develop faster, the mass of the primary particle can be estimated statistically in two ways. 
First, the atmospheric depth of the shower maximum, \Xmax, is on average smaller for heavier particles. 
\Xmax ~can be measured best by optical and radio detectors, where the most accurate measurements of energy and \Xmax ~are currently provided by fluorescence telescopes \cite{AugerComposition2017}, because the systematic uncertainties are well understood. 
Nevertheless, recent radio arrays have already reached similar accuracy \cite{LOFAR_Xmax,TunkaRex_Xmax,AERA_energyPRL2016}. 
Second, more muons and less electrons are produced compared to showers of light primary particles.
This can be used to estimate the primary mass by measuring the muonic and electromagnetic components of the same air shower separately, since it influences the ratio between the numbers of muons and electrons at the shower maximum as well as on ground.
\par 
An established method is to measure the number densities of electrons and muons at a reference distance to the shower axis on ground (e.g. in CASA-MIA \cite{Gibbs:1988cd,Glasmacher:1999xn}, AGASA \cite{Chiba:1991nf}, KASCADE-Grande \cite{Apel:2010zz}, and AugerPrime, the Upgrade of the Pierre Auger Observatory \cite{Aab:2016vlz}).
Except for showers more inclined than \degree{40}, the muons rarely interact or decay in the atmosphere and their number is approximately constant from the shower maximum to the ground.
However, the electrons are partly absorbed in the atmosphere and suffer larger energy losses so that their number depends on the distance to the shower maximum. 
Especially for showers at large zenith angles, the distance to the shower maximum is long and the number of electrons falls below the detection threshold.
On the contrary, the radio emission is produced by the electrons and positrons along the shower and is not absorbed in the atmosphere.
Thus, the radio signal provides information about the size of the electromagnetic component for all zenith angles.
Furthermore, the width of the radio footprint on ground rises with the zenith angle \cite{Aab:2018ytv}, which enables the economic detection of inclined showers with radio antenna arrays with a large spacing.
\par 
We studied the combination of radio and muon detection for the site of the the Pierre Auger Observatory \cite{ThePierreAuger:2015rma} in Malarg\"ue, Argentina, where the combination of muon and radio detectors is already realized.
Its main detector is dedicated to measure cosmic rays above $10^{18.5}$\,eV and comprises a $3000\,$km$^2$ large surface array of water-Cherenkov detectors overlooked by fluorescence detectors at four sites \cite{AugerFD2010}.
In the enhancements area of the Observatory, the AMIGA (Auger Muons and Infill for the Ground Array) \cite{PierreAuger:2016fvp} and AERA (Auger Engineering Radio Array) \cite{Holt:2017dyo} detectors measure showers in coincidence down to energies of $10^{17.5}$\,eV.
AMIGA consists of water-Cherenkov detectors on an area of 23\,km$^2$, arranged on a dense grid of 750\,m spacing compared to 1500\,m in the standard array, to lower the energy threshold.
Below the water-Cherenkov detectors, underground scintillators (AMIGA Muon Detector) are being installed at 2.3\,m depth as part of the AugerPrime Upgrade.
The water-Cherenkov detectors are sensitive to all shower particles whereas the underground scintillators are shielded from the electromagnetic component of the shower.
Thus, the underground detectors solely measure muons with an energy threshold of about $1\,$GeV.
AERA comprises antenna stations distributed on an area of 17\,km$^2$ inside AMIGA to detect the radio emission of cosmic-ray air showers in the frequency band of 30 --  80\,MHz.
By measuring the radio emission, AERA is mainly sensitive to the charged electromagnetic component of the shower.
Motivated, by the existing AMIGA and AERA arrays, we study for the same simulated air showers muons above $1\,$GeV and the radiation energy in the band of 30 --  80\,MHz.
\par 
In this work, the capability of the radio emission is evaluated to serve as a mass estimator in combination with the muons reaching ground.
To study the pure shower physics, air-shower simulations mostly independent of detector properties are used.
Apart from the energy threshold for muons and the radio frequency band, we do not simulate any specific detector response or array spacing, but instead study particle numbers and densities and the total radiation energies.
To facilitate the application to the real detectors, we use the geomagnetic field and height above sea level of the Auger site of \obslev.
Different parameters such as the particle numbers and radiation energy are studied for proton- and iron-induced air showers.
The mass separation power is investigated depending on the zenith angle and the primary energy.
It shows that in particular for inclined showers using the radio emission is superior compared to classical detection methods using solely particles on ground.

\section{Air-shower simulations}

The air-shower simulations used in this work are calculated with the simulation code CORSIKA \cite{Heck:1998vt}, using the hadronic interaction model QGSJETII-04 \cite{Ostapchenko:2010vb}.
The true particle distributions calculated from CORSIKA are used to investigate the muon density at different distances to the shower axis and the radiation energy.
To facilitate the foreseen application, the simulations used in this work were prepared in the energy range of the AMIGA Muon Detector and AERA. 
Thereby two simulation sets are used:
\begin{enumerate}
\item \begin{itemize} \item fixed zenith angle of $38^{\circ}$ \item energy range $10^{17.5} - 10^{19}$\,eV, isotropically distributed  \item azimuth \degree{0} -- \degree{360}, isotropically distributed \item 1000 simulations \end{itemize}
\item distribution of the direction and energy according to AERA measurements: \begin{itemize} \item energy range of $2 \cdot 10^{16} - 4 \cdot 10^{19}$\,eV \item full zenith and azimuth angle range \item 10604 simulations \end{itemize}
\end{enumerate}
Both sets contain an equal number of proton and iron primaries.
The air showers are simulated until an observation level of \obslev, corresponding to 870\,g\,$\cdot$\,cm$^{-2}$ atmospheric overburden and according to the average altitude of the AMIGA and AERA detectors.
The Earth's magnetic field is simulated according to the Auger site.
For all results, only muons above an energy of 1\,GeV are considered, corresponding to the energy threshold of the AMIGA Muon Detector at 2.3\,m underground, i.e. an additional overburden of 540\,g\,$\cdot$\,cm$^{-2}$ due to the soil \cite{PierreAuger:2016fvp,PhDHolt:2018}.
In order to study the general potential of combining muon and radio detection, we do not apply the specific detector responses or array layouts of AMIGA or AERA, but work with the true observables from the CORSIKA simulations.

\section{Definitions and methods}
\label{sec_methods}
In this work, the mass separation power is evaluated using the figure of merit ($FOM$).
The figure of merit quantifies the separation of two classes for an observable that has different mean values for each class, e.g, the seperation of proton and iron showers by the muon number.
The $FOM$ is defined as the following relation between the mean value and the standard deviation of the observable for the two classes:
\begin{equation}
FOM = \frac{ | \mu_{Fe} - \mu_p | }{ \sqrt{ \sigma_p^2 + \sigma_{Fe}^2 } } \quad ,
\end{equation}
where $\mu_{\textrm{i}}$ are the mean values and $\sigma_i$ the standard deviations of the proton (p) and iron (Fe) classes.
The figure of merit is a valid statistical estimator for Gaussian distributions.
For example, a $FOM = 1$ indicates that the means of the two classes are separated by one standard deviation of the difference between these observables.
In particular, we use the $FOM$ to quantify the separation between proton and iron showers based on the true observables derived from the CORSIKA simulations.
\par 
To calculate the mean and the standard deviation of the investigated observables in this study, they are normalized to an energy of $10^{18}$\,eV for all simulated air showers.
The energy dependencies are derived from power-law fits to the complete set of simulations:
\begin{equation}
a \left( E \right) = a_0 \cdot \left(\frac{E}{10^{18} \,\textrm{eV}} \right)^{\gamma} \quad ,
\end{equation} 
where $a(E)$ is any energy-dependent observable, $a_0$ its value at $10^{18}$\,eV, $E$ the primary energy used in the simulation, and $\gamma$ the index of the power law used to approximate the energy dependence.

\section{The muonic component}

\begin{figure}
  \includegraphics[width=0.485\textwidth]{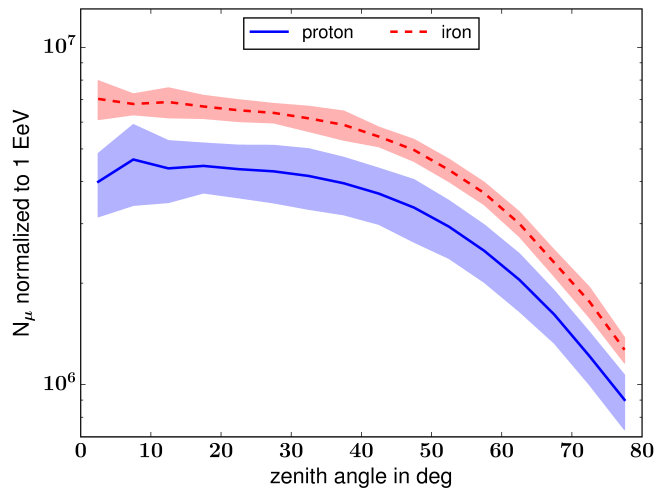}
\caption{Zenith angle dependence of the mean true number of muons at observation level (E$_{\mu} > 1$\,GeV). The number of muons is normalized to an energy of $10^{18}$\,eV. For zenith angles larger than \degree{40} the traveled distance becomes large and a fraction of the muons decays during the shower development before reaching the observation level of \obslev . As also in the following figures, the lines and their surrounding bands denote the mean values and the standard deviations.} 
\label{fig:Muonsground}
\end{figure}

\begin{figure*}
	\subfloat[proton-induced showers][]{
		\includegraphics[height=6.5cm]{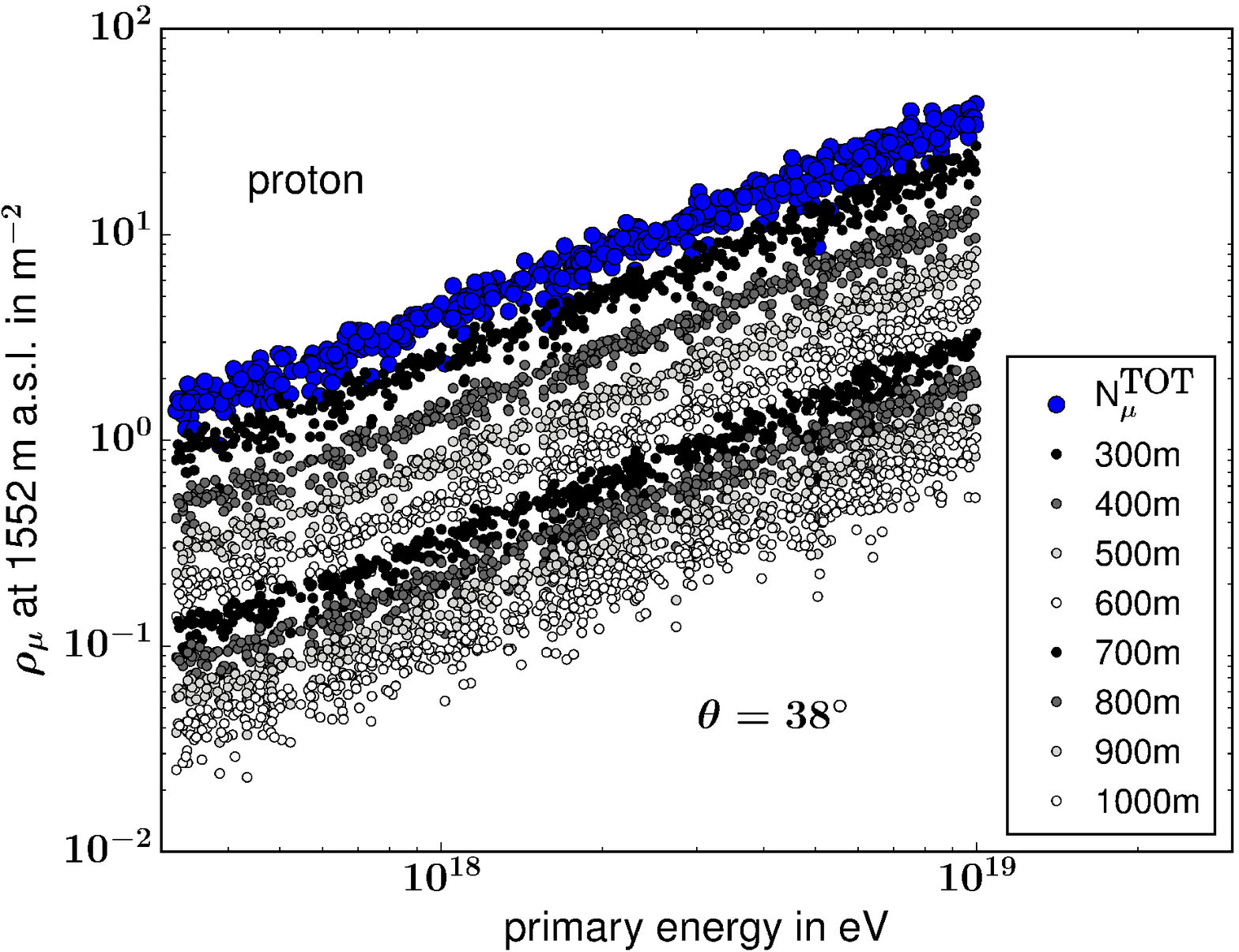}
		\label{fig:muonsDensEP}
	}
	\hspace{-0.3cm}
	\subfloat[iron-induced showers][]{
		\includegraphics[height=6.5cm]{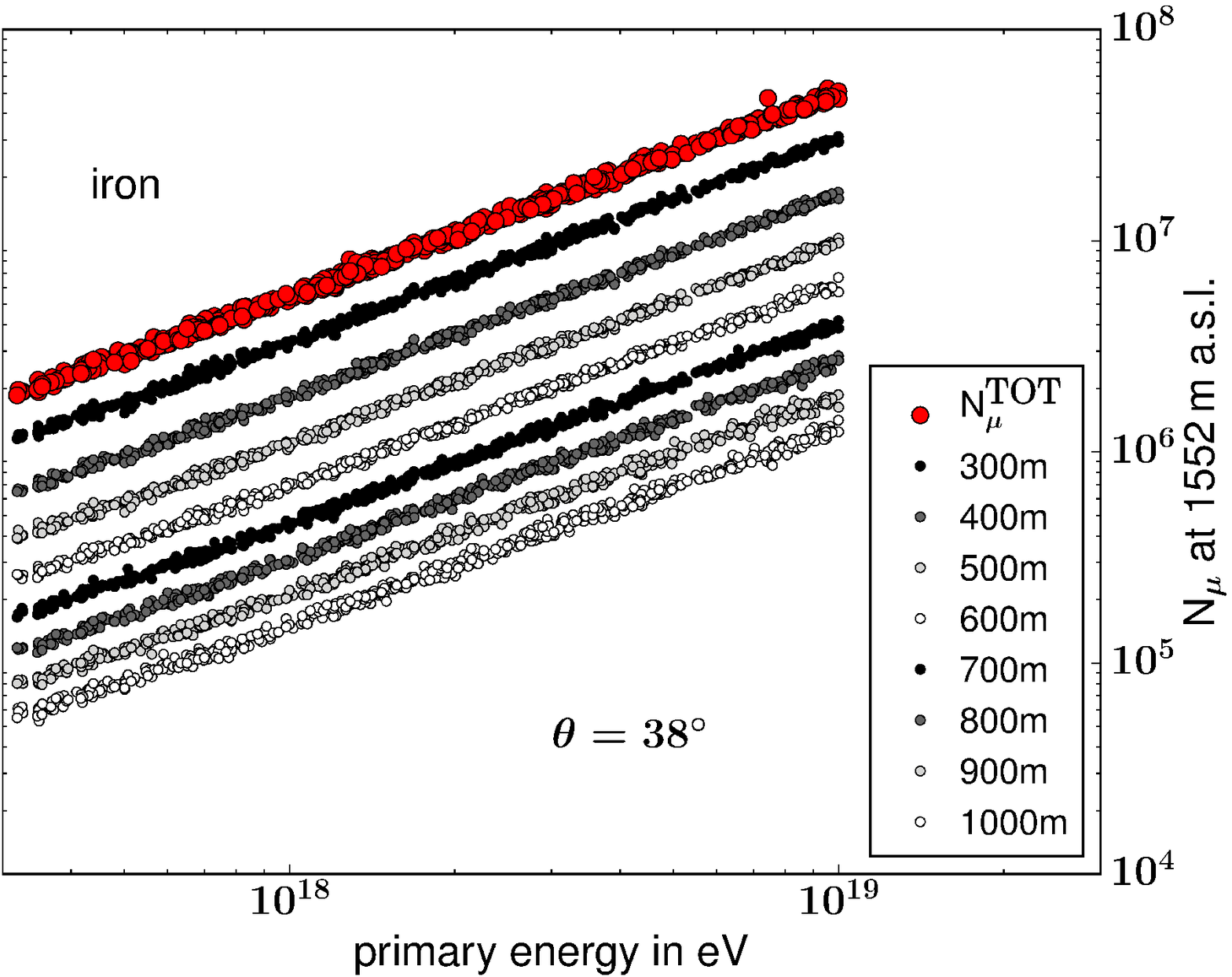}
		\label{fig:muonsDensEFe}
	}
	\caption{True muon density at different distances to the shower axis above a muon energy of 1\,GeV for showers induced by a proton (a) and an iron nucleus (b) with a zenith angle of \degree{38}. The muon density is directly correlated to the total number of muons on ground and shows approximately the same energy dependence.}\label{fig:muonsDensE}
\end{figure*}

Muons interact much less with matter, i.e.~the atmosphere, than electrons and have only negligible energy losses from bremsstrahlung due to their larger mass.
The muon is an unstable particle, but its decay is only mediated by the weak interaction and therefore slow with a mean lifetime of 2.2\,$\upmu$s.
Hence, this decay only becomes important for inclined showers, where the distance from \XmaxSpace to the observation level is of the same order of the distance traveled in the (time dilated) lifetime.
In Ref.~\cite{PhDHolt:2018} it is shown, that the number of muons at \XmaxSpace and on ground deviates for zenith angles above \degree{40}.
Moreover, the total number of muons of a shower when reaching the ground depends strongly on the primary energy.
A power-law fit to the simulations with the full zenith angle range shows that the muon number increases with the energy with a mean index of $\gamma = 0.93$ (for proton $\gamma = 0.922 \pm 0.004$ and for iron $\gamma = 0.934 \pm 0.004$).
This mean index will be used in the following to correct for the energy dependence of the number of muons $N_\mu$.
For comparison, older versions of the different hadronic models predict similar values between $\gamma = 0.88 - 0.92$ \cite{AlvarezMuniz:2002ne}.
\par 
In Fig.~\ref{fig:Muonsground} the mean true number of muons $N_\mu$ at an observation level of \obslev is plotted over the zenith angle.
For each shower the number of muons is normalized to the corresponding value at an energy of $10^{18}$\,eV, using the obtained fit results of the energy dependence.
For zenith angles up to around \degree{40} the number of muons is stable up to the observation level.
For larger zenith angles a growing fraction of the muons decays before reaching the observation level. 
\par  
The number of muons at the observation level is larger for iron than for proton showers for all zenith angles.
The spread (standard deviation) is larger for proton showers due to larger shower-to-shower fluctuations. 
However, the separation between proton and iron is larger than the spread, which shows the potential of the number of muons on ground for estimation of the primary mass, provided that the primary energy is known.


\subsection{Observable: muon density at a reference distance}\label{sec:trueMuonDens}

The particle footprint of a cosmic-ray air shower extends over several square kilometers on ground at the energies investigated here.
Therefore, realistic experiments cannot measure measure the total number of muons directly, but instead locally measure the number of muons at several positions with sparse detector arrays. 
This corresponds to a measurement of the muon density $\rho_{\mu}$ (= number of muons per area) at certain distances from the shower axis. 
Thus, we calculated the muon density at certain distances from the shower axis from the muon output of the CORSIKA simulations.
As shown in Fig.~\ref{fig:muonsDensE}, the muon density at a chosen distance is directly correlated to the total number of muons at the observation level and can be used as an observable for the latter.
The figure shows the true muon density for proton (Fig.~\ref{fig:muonsDensEP}) and iron (Fig.~\ref{fig:muonsDensEFe}) primaries at distances from 300 -- 1000\,m as well as the true total number of muons at an observation level of \obslev compared to the primary energy for showers with a zenith angle of \degree{38}.
The muon density decreases with the distance to the shower axis.
Furthermore, the muon density shows a slightly lower energy dependence (smaller index $\gamma$) than the total number of muons for all distances, e.g., for a distance of 600\,m in the simulations including all zenith angles the mean index is $\gamma = 0.90$ (for proton $\gamma = 0.894 \pm 0.002$ and for iron $\gamma = 0.907 \pm 0.002$).
As explained below, we have selected 600\,m as reference distance because of the high mass separation power of the muon density at 600\,m.
Hence, in the following the muon density is normalized to a primary energy of $10^{18}$\,eV with an index of $\gamma = 0.90$.
\par
\begin{figure}
  \includegraphics[width=0.485\textwidth]{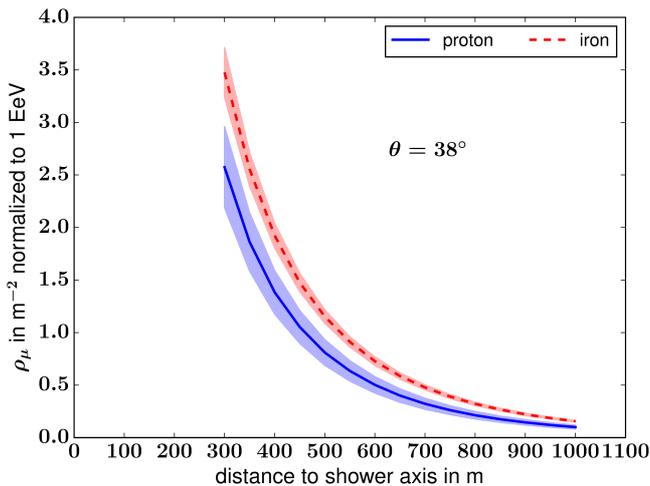}
\caption{True muon density compared to the distance to the shower axis for showers with a zenith angle of \degree{38}, normalized to $10^{18}$\,eV. The muon density decreases with the distance. Thereby, the relative difference between proton- and iron-induced showers increases, whereas the relative spread is constant.}
\label{fig:MuonDensLDF1GeV}
\end{figure}

\begin{figure}
  \includegraphics[width=0.485\textwidth]{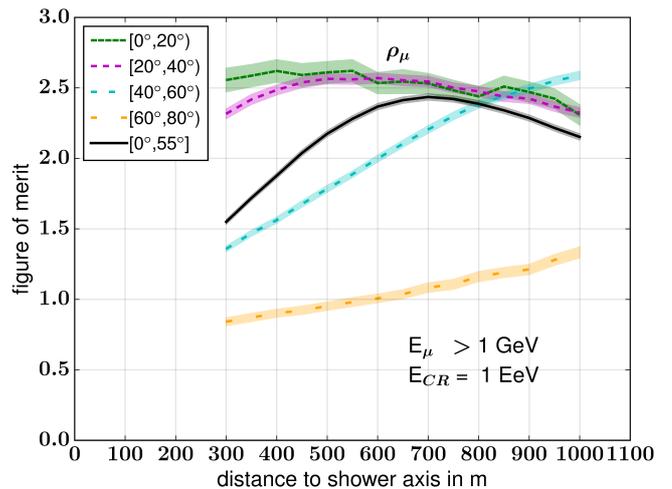}
\caption{Figure of merit compared to the distance to the shower axis for different zenith angle ranges and for combining showers at all zenith angles. The muon density is normalized to a primary energy of $10^{18}$\,eV for all showers. The figure of merit decreases with the zenith angle due to larger shower-to-shower fluctuations. It shows a maximum between 600 -- 800\,m for the range of \degree{0} -- \degree{55}.}
\label{fig:MuonDensFOM1GeV}
\end{figure}

 
The mean muon density is shown over the distances of 300\,m -- 1000\,m from the shower axis in Fig.~\ref{fig:MuonDensLDF1GeV} for showers with a zenith angle of \degree{38}, normalized to a primary energy of $10^{18}$\,eV.
The muon density decreases with the distance to the shower axis for both proton and iron showers.
The mass separation power depends on the relative difference and the spread at each distance, which is quantified by the figure or merit shown for different ranges of zenith angles in Fig.~\ref{fig:MuonDensFOM1GeV}.
In addition, the figure of merit is shown for the zenith angle range of \degree{0} -- \degree{55}, at which AMIGA features full detection efficiency \cite{PhDSchulz:2016}. 
For zenith angles below \degree{20}, the figure of merit slightly decreases with the distance.
For showers more inclined than \degree{40}, it increases with the distance and does not reach a maximum until 1000\,m.
Combining the zenith angles from \degree{0} to \degree{55} leads to a broad maximum in the figure of merit between 600\,m and 800\,m for a muon energy threshold of 1\,GeV.
Since we expect real experiments to suffer less from measurement uncertainties at higher values of $\rho_\mu$, we have selected 600\,m as reference distance for the present study.
\par 

\begin{figure}
  \includegraphics[width=0.485\textwidth]{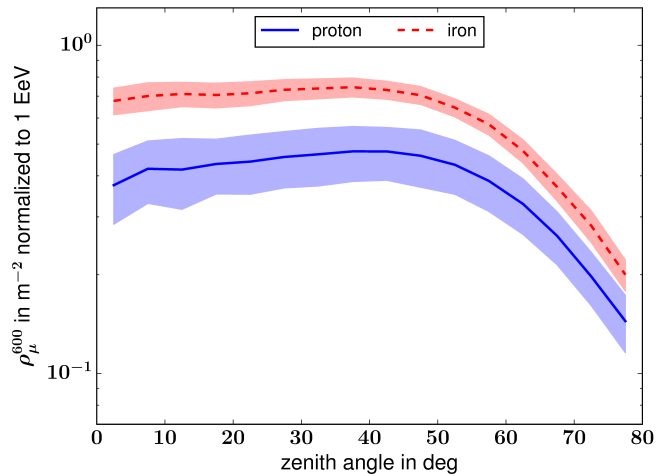}
\caption{Zenith angle dependence of the muon density at 600\,m above a muon energy of 1\,GeV. The muon density increases slightly for zenith angles up to \degree{50}, which indicates that up to these zenith angles the production of high energy muons dominates. For larger zenith angles it decreases due to muon decay.}
\label{fig:MuonDensZen1GeV}
\end{figure}

The dependence of the muon density \Rhomu on the zenith angle is shown in Fig.~\ref{fig:MuonDensZen1GeV}.
It slightly increases for zenith angles up to around \degree{50} and decreases again at higher zenith angles due to muon decay.

\section{The electromagnetic component and the radio emission}

\begin{figure}
  \includegraphics[width=0.485\textwidth]{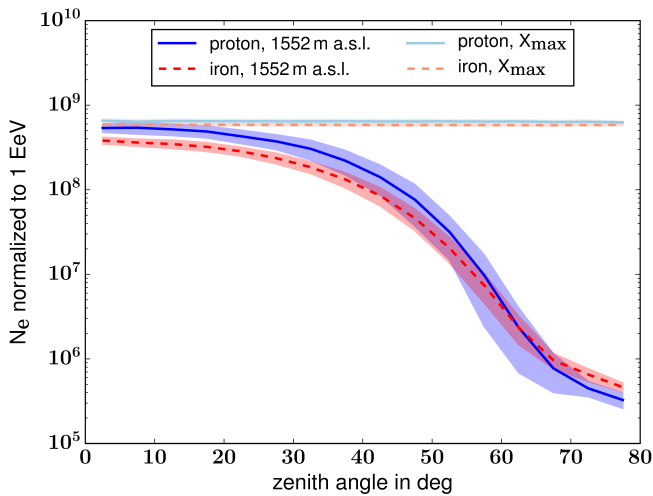}
\caption{Zenith angle dependence of the true number of electrons above an energy of 250\,keV. The number of electrons at \obslev decreases by about three orders of magnitude over the plotted zenith angle range due to absorption in the atmosphere. Proton showers contain more electrons than iron showers except for very inclined showers, where the electrons are mainly products of muon decays. On the contrary, the number of electrons at \XmaxSpace is nearly independent of the zenith angle and higher in proton showers.}
\label{fig:NeGroundXmax}
\end{figure}

The number of electrons $N_e$ in an air shower can be measured by particle detectors on ground in a similar way as the muons.
However, the electrons are partly absorbed in the atmosphere on their way to the ground and suffer much larger energy losses, e.g. by brems\-strah\-lung, than the much heavier muons.
Therefore, their number in the shower strongly depends on the distance in atmospheric depth to \Xmax, which increases with the zenith angle $\theta$ roughly with $1/\cos{\theta}$.
The number of electrons at an observation level of \obslev and at the electromagnetic \XmaxSpace are plotted over the zenith angle in Fig.~\ref{fig:NeGroundXmax}.
As expected, the number at the observation level decreases with the zenith angle and is about three orders of magnitude smaller at an angle of \degree{80} compared to vertical showers.
This dependence on the zenith angle has to be taken into account when using the electron number for mass separation measurements, which leads to additional uncertainties from the measurement of the arrival direction of the shower. 
Depending on the size and type of the particle detectors and on the observation level, the number of electrons falls below the detection threshold and only the muons are detected for very inclined showers.

Moreover, it becomes apparent that the difference between proton and iron showers becomes smaller and finally flips the direction, so that for $\theta >$ \degree{65} iron showers contain more electrons at the observation level than proton showers do.
The shower-to-shower fluctuations are increased around the zenith angle of the flip.
In addition, the slope becomes smaller towards higher zenith angles.
These observations are explained by the increasing number of muon which decay into electrons (cf. Fig.~\ref{fig:Muonsground}). 
Hence, these electrons are mostly created by the muonic component.
Thus, for inclined showers, the number of electrons is correlated with the number of muons in the shower, which is larger for heavier primary particles. 
Due to this flip in the proton-iron separation and the strong decrease, the number of electrons at the observation level does not provide a reliable mass estimator for inclined air showers.
\par 
On the contrary, the number of electrons at \XmaxSpace does not depend on the zenith angle.
Independent of the shower inclination, it is larger for proton than for iron-induced air showers and, thus, provides information about the mass of the primary particle.
However, the number of electrons at \XmaxSpace cannot be measured directly by air-shower arrays on ground, but indirectly by the electromagnetic energy deposited in the atmosphere.
This electromagnetic shower energy is slightly larger for proton than for iron showers, e.g., by 4.5\,\% for a primary energy of $10^{18}$\,eV and 3\,\% at $10^{19}$\,eV \cite{Engel:2011zzb}. 
The electromagnetic energy of an air shower can be measured by its radio emission and in dark clear nights additionally by the fluorescence light and the Cherenkov light produced in the atmosphere.
Hence, in contrast to direct measurements of the number of electrons on ground, indirect measurements of the number of electrons at \XmaxSpace by radio (or optical) detectors provide a useful observable for the combination with muon measurements over all zenith angles including very inclined showers.


\subsection{Observable: the radiation energy of the radio emission}

The radio emission of an air-shower is induced by the electromagnetic particles in the shower \cite{Schroder:2016hrv,Huege:2016veh}.
Hence, the energy contained in the radio emission - the radiation energy $E_{\textrm{rad}}$ - provides a calorimetric measurement of the electromagnetic shower energy $E_{\textrm{em}}$. 
This correlation between the radiation energy and the electromagnetic shower energy was observed at the Pierre Auger Observatory \cite{Aab:2015vta}.
It was modeled and corrected for various dependencies on the arrival direction in Ref.~\cite{Glaser:2016qso}, which is used here to calculate a corrected radiation energy \Srd from the shower simulations (see \ref{app:A}). 
Thereby, the radio emission in the frequency band of 30 -- 80\,MHz is considered. 
\par 

\begin{figure}
	\includegraphics[width=0.485\textwidth]{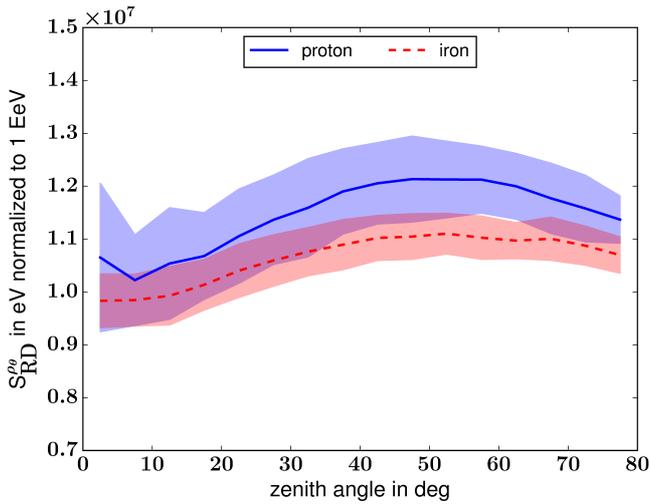}
	\caption{True radiation energy compared to the zenith angle. The radiation energy is corrected for the angle between the geomagnetic field and the shower axis as well as for the air density at the mean \XmaxSpace at the corresponding zenith angle. Furthermore, the radiation energy is calculated for the observation level of \obslev. The radiation energy increases with the zenith angle up to \degree{50}, whereas this effect is larger for proton showers. This leads to an enhanced mass sensitivity at this zenith angle.}
	\label{fig:Srd}
\end{figure}

The corrected radiation energy \Srd is plotted over the zenith angle in Fig.~\ref{fig:Srd} after normalization to $10^{18}$\,eV by assuming \Srd $\propto E^2$ according to Ref.~\cite{Aab:2015vta}.
\Srd grows with the zenith angle, as inclined showers extend over a larger geometric distance on which the radiation energy is released.
In addition, the shower-to-shower fluctuations decrease with increasing zenith angle.
Proton showers release more radiation energy than iron showers due to the larger amount of electrons and positrons at \Xmax.
In fact, the difference between proton and iron showers grows with the zenith angle.
This is expected, since the mean free path of the shower particles grows and, thus, the difference between the total path lengths of all electrons and positrons.
In addition, the full development of an iron shower is shorter in atmospheric depth than of a proton shower with the same primary energy.
This becomes apparent, when the iron shower (A=56) is described as 56 parallel developing "proton" sub-showers with each a primary energy of E\textsubscript{0}/56 as in Ref.~\cite{Matthews:2005sd}.
Each of these sub-showers develops faster than the proton shower with the primary energy E\textsubscript{0} and hence the whole iron shower does.
Therefore, the proton shower travels a longer geometric distance on which more radiation energy is released.
This effect becomes larger for more inclined showers, where the ratio between the atmospheric depth and the geometric distance becomes larger.



\section{Mass estimation by combining observables}\label{sec:trueMassSepComb}
Combining the two observables, the muon density and the radiation energy, leads to a mass sensitive parameter introduced in this section.
Summarizing the results for the electromagnetic component, the radiation energy shows an enlarged mass sensitivity at higher zenith angles, reaching a plateau at around \degree{50}.
In contrast, the number of electrons looses its mass sensitivity at angles above \degree{60} due to the fact that mainly electrons originating from muon decay reach the observation level.
In addition, the uncertainties due to shower-to-shower fluctuations are particularly large for inclined showers.
The muonic component shows a different behavior.
The difference between proton and iron showers for the muon density at 600\,m only decreases slowly for large zenith angles.

\begin{figure}
	\includegraphics[width=0.485\textwidth]{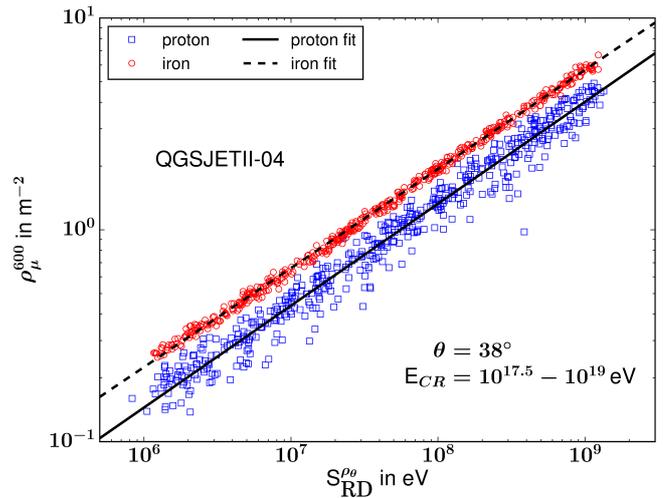}
	\caption{Correlation between the muon density and the radiation energy and a power-law fit.}
	\label{fig:RhomuoverSrd}
\end{figure}

\begin{figure}
	\includegraphics[width=0.485\textwidth]{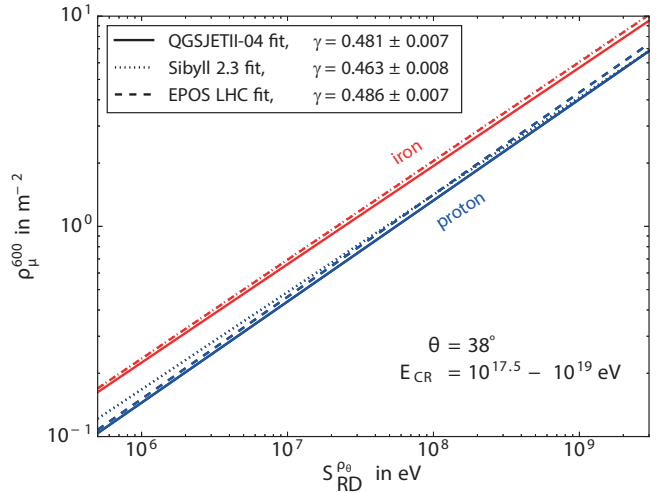}
	\caption{The correlation between the muon density and the radiation energy (as in Fig.~\ref{fig:RhomuoverSrd}) using different hadronic models. The average index $\gamma$ of a power-law fit to the proton and iron distributions shows only minor differences. Therefore, only QGSJETII-04 is used for all following investigations.}
	\label{fig:RhomuoverSrdModels}
\end{figure}

The quantities of the electromagnetic component feature higher values for proton showers ($N_e$ only up to \degree{60} zenith angle), the muon density is higher for iron.
Therefore, it is expected that the mass sensitivity is enhanced by combining these complementary observables.
The muon density is plotted over the radiation energy in Fig.~\ref{fig:RhomuoverSrd} for QGSJETII-04 simulations with a zenith angle of 38$^{\circ}$.
The simulations are repeated using different hadronic interaction models, i.e. Sibyll 2.3 \cite{Riehn:2017} and EPOS-LHC \cite{Pierog:2013ria}, for comparison in Fig.~\ref{fig:RhomuoverSrdModels} (shown as power-law fits to the simulations). 
Sibyll 2.3 predicts more muons for proton showers at small radiation energies and EPOS-LHC at higher radiation energies.
The models mainly differ in the absolute scale of the predicted muon density. 
The absolute scale is relevant for the interpretation of a mass estimator in terms of atomic mass numbers, but not for the separation between light and heavy primaries investigated here. 
The difference in the indices of the power law used to normalize the energy dependence (cf.~Sec.~\ref{sec_methods}) is statistically significant, but small in size. 
Therefore, we have not examined the detailed impact of using different hadronic interaction models in the context of this conceptual study, and performed the full analysis using QGSJETII-04.

\begin{figure}
	\includegraphics[width=0.485\textwidth]{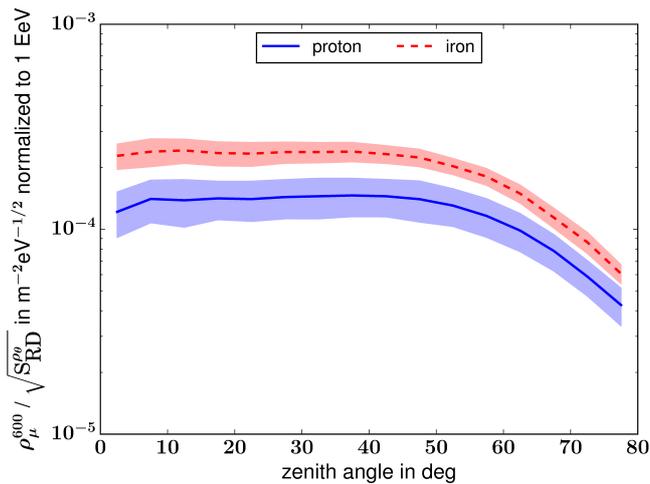}
	\caption{Ratio between the muon density and the square root of the radiation energy compared to the zenith angle. The separation of proton and iron showers exceeds the spread originating from shower-to-shower fluctuations. The ratio decreases for larger zenith angles, since the muon density decreases (cf. Fig.~\ref{fig:Muonsground}).  }
	\label{fig:RatioRhomuSrd}
\end{figure}

In Fig.~\ref{fig:RatioRhomuSrd} the ratio between the muon density at 600\,m axis distance and the square root of the radiation energy is plotted over the zenith angle.
Fitting the ratio over the primary energy with a power law for the simulations over all zenith angles leads to a correlation of \RhomuSrd $\propto$ E$^{0.058}$, which is used here to normalize the ratio to $10^{18}$\,eV.
Since both, the muon density at 600 m and the radiation energy, slightly increase with zenith angle until about \degree{50}, the ratio is nearly independent of the shower inclination until \degree{50} zenith angle. 
It decreases at zenith angles above \degree{50}, since the muon density decreases (see Fig.~\ref{fig:Muonsground}).
The separation between proton and iron showers is larger than the spreads (standard deviations) of both distributions for all investigated zenith angles.

\begin{figure*}
	\includegraphics[width=\textwidth]{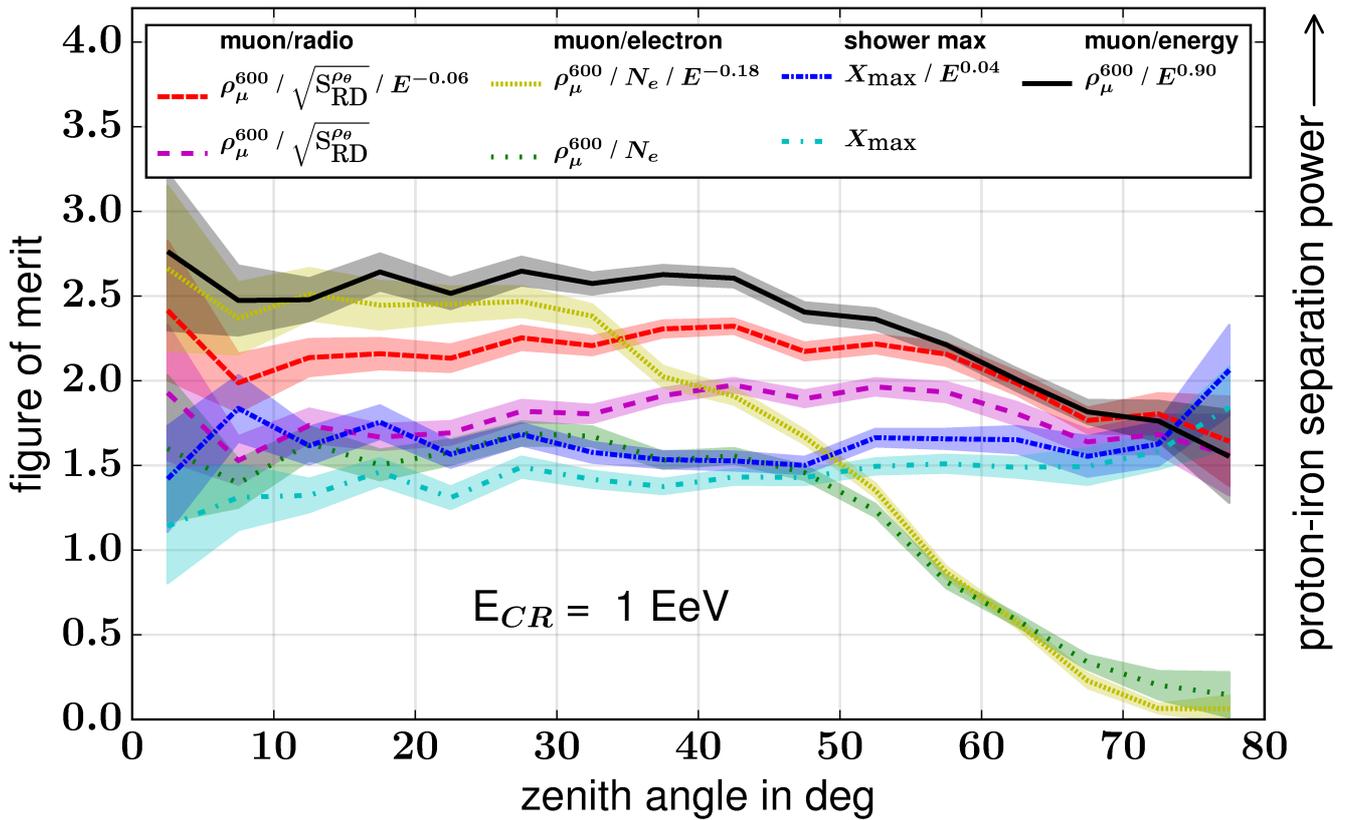}
	\caption{Figure of merit of the different shower observables. The different mass estimators are corrected for their dependence on the true primary energy, which is known as input parameter of each simulation. In addition, the uncorrected ratios are shown providing a more realistic estimate of the potential for real air-shower arrays. 
The bands depict the uncertainties due to shower-to-shower fluctuations. 
All observables are true values derived from CORSIKA simulations.
They do neither include the effects of a specific layout of an air-shower array nor detector specific uncertainties.}
	\label{fig:FOMall}
\end{figure*}

The mass separation power represented by the figure of merit is shown in Fig.~\ref{fig:FOMall} for the ratio of the muon density and the radiation energy, the ratio of the muon density and the number of electrons at \obslev, the muon density, and \Xmax. 
The figure of merit of \Xmax ~serves as a reference, since \Xmax ~is a widely used observable for measurements of the mass composition \cite{Kampert:2012}.
The \Xmax ~values are directly obtained from the same CORSIKA simulations as the other observables, not including any assumptions on a specific detection technique. 
All observables are shown with and without normalization for their dependencies on the true energy of the primary particle, which is known exactly in the simulations.
The muon density alone has only mass separation power if the primary energy is known, and therefore has no counterpart without normalization. 
In a realistic experiment, the primary energy might not be reconstructed accurately enough for the purpose of normalization. 
Therefore, we also show the figure of merit for the unnormalized observables, so the potential improvement by normalization for the primary energy can be assessed by comparing both sets of curves. 
Whether or not normalizing for the primary energy, the classical method of the muon-electron ratio looses mass sensitivity with increasing shower inclination. 
In contrast, both \Xmax ~and the new combination of the muon density and the radiation energy can be used for the purpose of mass separation for the full range of investigated zenith angles.

\section{Discussion}
The results shown here compare the well established mass estimator using the particles numbers with the novel method combining the muons with the radio emission.
The first proof of principle conducted here illustrates the potential of radio-emission measurements to enhance mass estimation in particular for inclined showers.
Potentially, the mass sensitivity can be improved further by investigating other radio observables, such as the radio amplitude or energy fluence at a reference distance instead of the integral \sqrtSrd over the whole footprint, and by tuning the reference distance for the muon density as a function of zenith angle instead of using a fixed distance of 600\,m.
Furthermore, the method of mass estimation via the ratio of the muon density and the radiation energy can be combined with the independent mass estimator \Xmax, measurable by radio as well as optical detectors, to reduce the overall uncertainties on the mass.
\par
Fig.~\ref{fig:FOMall} shows the intrinsic mass sensitivities of various observables not including detector effects and measurement uncertainties.
In addition to the uncertainties of the individual observables, the uncertainty on the reconstructed energy of the primary particle will impact the total accuracy for the mass. 
Therefore, it is an advantage of the new radio-muon mass estimator that it only weakly depends on the energy of the primary particle.
Compared to the electron-muon ratio, the normalization for the energy has a relatively small influence on the figures of merit for the radio-muon combination and for \Xmax.
In particular \XmaxSpace and the energy content of the electromagnetic shower component can be measured not only by radio arrays but also by other techniques. 
Due to their similarities in the sensitivity to the electromagnetic shower component, qualitatively similarly results can be expected for the combination of muon detection with fluorescence or air-Cherenkov light.
However, these techniques suffer from their limited duty cycle and atmospheric light absorption. 
The latter hampers the air-Cherenkov measurement particularly for inclined showers.
Hence, only the combination of muon detectors with either fluorescence or radio detectors is expected to provide high mass sensitivity for large zenith angles.
Coincident events with the fluorescence, muon, and radio detectors of the upgraded Pierre Auger Observatory will enable an independent cross-check of the new mass estimator. 
\par 
The influences of realistic detector responses, Poissionian fluctuations due to limited detector sizes, measurement uncertainties, and background is investigated for the combination of the AMIGA Muon Detector and AERA of the Pierre Auger Observatory in Refs.~\cite{PhDHolt:2018,Holt:2018a}.
As expected, these effects slightly degrade the mass-separation power, but generally the high potential for mass-composition studies is confirmed.
Dedicated simulation studies need to be done to estimate the full potential of the radio-muon combination for showers more inclined than \degree{55}, since the reference distance in this study was optimized for the range of zenith angles until \degree{55} accessible by AMIGA.
\par
As for other methods for the estimation of the mass of the primary particle based on air-shower observable, the use of hadronic interaction models comes with unavoidable systematic uncertainties. 
As studied by several experiments \cite{DembinskiUHECR2018}, all available hadronic interactions models have a deficit in the prediction of muons, which is largest at the highest energies \cite{Aab:2016hkv}.
We do not expect that this discrepancy in the muon number will have a significant influence on the mass separation power investigated in this work, which relies on relative differences in the muon content of proton and iron showers. 
However, the absolute scale of the muon density is shifted, that has to be taken into account when comparing simulations to measured data and when interpreting data based on simulations.
\par 
Finally, the novel technique for mass estimation can be applied to other experiments.
While dedicated simulation studies will be needed for each experiment, the general findings of this study should be transferable, since neither high-energy muons nor the radio signal suffer from significant absorption in the atmosphere. 
Only for sites at higher altitudes the effect of partly clipping the air shower at the observation level needs to be investigated more carefully, in particular for vertical and mildly inclined showers.
\par
A variety of activities is already ongoing. 
The Pierre Auger Observatory is currently being upgraded with scintillators on top of the water-Cherenkov detectors to disentangle the muonic and electromagnetic components of showers up to zenith angles of \degree{60}.
Investigations are ongoing to equip each surface detector station in addition with a radio antenna, which will allow to measure the electromagnetic component as well for inclined shower \cite{Huege:2015lga,Horandel:2018a}.
Possibilities to lower the energy threshold of the radio detection technique by the right choice of the frequency band were investigated in Ref.~\cite{V.:2017kbm}.
This can be applied to search for air showers induced by PeV photons using the radio-muon combination for gamma-hadron separation.
Therefore, activities are ongoing to enhance the IceTop particle detector array at the South Pole with radio antennas \cite{Schroder:2018a}.
Furthermore, the TAIGA facility comprises muon detectors and radio antennas (Tunka-Rex), at which the technique could directly be applied \cite{Budnev:2017fyg}. 
The Giant Radio Array for Neutrino Detection (GRAND) is a huge antenna array planned in China focusing on inclined showers \cite{Alvarez-Muniz:2018bhp}.
GRANDproto300, its next stage prototype, will be additionally equipped with Auger-like particle detectors for the purpose of applying the radio-muon method to cosmic-ray air showers.

\section{Conclusion}

In this work, a novel technique is developed to estimate the mass of ultra-high energy cosmic rays by combining the muon signal and the radio emission of air showers.
The muonic and electromagnetic components of air showers induced by proton and iron primaries were analyzed based on air-shower simulations.
The size of the muonic component can be observed by the muon density $\rho_{\mu}(\textrm{r}_{\textrm{ref}})$ at a reference distance to the shower axis.
600\,m was found to be a distance with a strong mass sensitivity for showers with zenith angles below \degree{55}, and the mass sensitivity may further be enhanced in particular for more inclined showers by varying the reference distance as a function of zenith angle.
The radiation energy \Srd, i.e. the energy contained in the radio emission, is correlated to the size of the electromagnetic component.
The ratio \RhomuSrd represents a mass-sensitive parameter that is larger for iron than for proton showers.
The mass sensitivity of the radio-muon combination was investigated and compared to established methods using solely particle measurements, or using the shower maximum \Xmax.
With the presented approach, the radio-muon combination features a mass separation power slightly larger than that of \XmaxSpace for all zenith angles. 
Only if the energy of the primary particle is known accurately enough, the traditional observable of the electron-muon ratio provides better mass separation for near-vertical showers with zenith angles smaller than  \degree{35}. 
Otherwise, and generally for more inclined showers, the new method of the radio-muon combination features 
superior mass estimation.

This emphasizes the potential for this new mass-sensitive parameter in particular for inclined showers.
At large zenith angles, the radio emission spans over a large area on ground, which makes sparse detection array and thus large-scale applications feasible.
The results show that the novel technique provides additional accuracy for mass measurements of cosmic rays on a per-event level.
This is essential for further progress in answering open questions about the origin of ultra-high energy cosmic rays, since not only the total flux, but also the mass composition is expected to feature an anisotropy in the arrival directions \cite{Aab:2016vlz}.
Therefore, the scientific potential of existing particle-detector arrays can easily be enhanced by adding radio antennas, and future air-shower arrays can be planned to feature both, muon and radio detectors.

\begin{acknowledgements}
This work was financed by the Helm\-holtz International Research Group of the Helm\-holtz Association, the Bundesministerium f\"ur Bildung und Forschung, Verbundforschung Auger, the Deutsch-Argentinisches Hoch\-schul\-zen\-trum/Centro Universitario Argentino-Alem\'{a}n (DAHZ-CUAA) and the Karlsruhe School of Elementary Particle and Astroparticle Physics: Science and Technology (KSETA). We would like to acknowledge the support of the Pierre Auger Collaboration, in particular from the colleagues of the AERA and AMIGA groups. We especially thank Tim Huege for very useful inputs.
\end{acknowledgements}

\appendix

\section{Calculation of the radiation energy from the electromagnetic shower energy}
\label{app:A}

The radiation energy is correlated to the electromagnetic shower energy as shown in Ref.~\cite{Aab:2015vta}.
The correlation was modeled in \cite{Glaser:2016qso} based on CoREAS simulations \cite{Huege:2013vt} for showers with zenith angles up to \degree{80}, including various corrections for dependencies on the arrival direction.
The model is used in this work to calculate the radiation energy from the shower energy extracted from the CORSIKA simulations.
\par 
The radiation energy slightly depends on the arrival direction and the angle $\alpha$ to the geomagnetic field.
The geomagnetic fraction of the radiation energy is influenced by the magnitude of the geomagnetic field $B_{\textrm{Earth}}$ as well as the angle $\alpha$ between the shower axis and $B_{\textrm{Earth}}$.
The radiation energy scales with $\sin^2{\alpha}$ because of the coherent nature of the radio emission.
The charge excess fraction $a$ of the radiation energy grows with the atmospheric density $\rho_{\textrm{\Xmax}}$ at \Xmax.
The atmospheric density decreases with altitude.
Thus, $\rho_{\textrm{\Xmax}}$ depends on the zenith angle and the altitude of the shower maximum \Xmax, which is generally higher in the atmosphere for showers induced by heavier particles. 
Furthermore, there is a second order dependence on $\rho_{\textrm{\Xmax}}$.
Whereas the radio emission depends on the geometric distance (in m), the air shower develops according to the atmospheric depth (in g/cm$^2$). 
The ratio between the geometric distance and the atmospheric depth is higher for regions of lower atmospheric density.
This leads to a slightly larger radio emission, if \XmaxSpace is higher in the atmosphere.
Therefore, the radiation energy $E_{\textrm{rad}}$ is corrected for these dependencies by
\begin{multline}
S_{\textrm{RD}}^{\rho_{\textrm{\Xmax}}} =  \frac{E_{\textrm{rad}}}{a(\rho_{\textrm{\Xmax}})^2 + \left( 1 - a(\rho_{\textrm{\Xmax}})^2 \right) \cdot \sin^2{\alpha}} \\ \cdot \frac{1}{\left( 1 - p_0 + p_0 \cdot \exp{ \left[ p_1 \cdot \left( \rho_{\textrm{\Xmax}} - \left< \rho \right> \right) \right] } \right)^2 } \quad ,
\label{eq:Erad}
\end{multline}
where $p_0 = 0.251 \pm 0.006$ and $p_1 = -2.95 \pm 0.06$\,m$^3$/kg, and $\left< \rho \right> = 0.65$\,kg/m$^3$ is the atmospheric density at the average $\left< \mathit{X}_\mathrm{max} \right>$ = 669\,g/cm$^2$ for an average zenith angle of \degree{45} and a primary energy of $10^{18}$\,eV for a 50\%-proton / 50\%-iron composition.
\par 
The correlation of the true radiation energy S\textsubscript{RD}, after normalization according to the arrival direction, with the electromagnetic shower energy $E_{\textrm{em}}$ is modeled in \cite{Glaser:2016qso} by
\begin{equation}
S_{\textrm{RD}}^{\rho_{\XmaxMath}} = A \cdot 10^7\,\textrm{eV} \cdot \left( \frac{E_{\textrm{em}}}{10^{18}\,\textrm{eV}} \right)^B
\label{eq:Srd}
\end{equation}
with $A = 1.683 \pm 0.004$ and $B = 2.006 \pm 0.001$.
The radiation energy used for this model was later found to be underestimated by about 11\% in the simulations used for this parametrization due to settings in the CoREAS simulations \cite{Gottowik:2017wio}.
Since this only affects the absolute scale, but not the relative difference between proton and iron showers, this is not considered here.
\par 
\XmaxSpace, which is used in the correlation, is often not accessible in an experiment and in case it is measured, there are additional measurement uncertainties.
Therefore, a correction dependent on the zenith angle, for which the measurement uncertainties are much smaller, is formulated based on the mean \XmaxSpace at the respective zenith angles:
\begin{multline}
S_{\textrm{RD}}^{\rho_{\theta} \prime} = \frac{E_{\textrm{rad}}}{a(\rho_{\theta})^2 + \left( 1 - a(\rho_{\theta})^2 \right) \cdot \sin^2{\alpha}} \\ \cdot \frac{1}{\left( 1 - p_0 + p_0 \cdot \exp{ \left[ p_1 \cdot \left( \rho_{\theta} - \left< \rho \right> \right) \right] } \right)^2 } \quad .
\label{eq:SrdZenUnclipped}
\end{multline}
In addition, another zenith angle dependent effect has to be taken into account.
Depending on the observation level of a detector, the shower might not be fully developed at the altitude of detection.
Hence, a part of the shower is clipped before the radio emission of this part is released.
The magnitude of this clipping effect depends on the distance between the observer and \Xmax.
The radiation energy investigated in the following is corrected for this effect by
\begin{equation}
S_{\textrm{RD}}^{\rho_{\theta}} = \frac{S_{\textrm{RD}}^{\rho_{\theta} \prime}}{1 - \exp{ \left(-8.7\,\textrm{cm}^2 / \textrm{kg} \left( D_{\textrm{\Xmax}} + 0.29\,\textrm{kg} / \textrm{cm}^{2} \right)^{1.89} \right) }}
\label{eq:SrdZen}
\end{equation}
where $D_{\textrm{\Xmax}}$ is the distance between the observer and \XmaxSpace in kg/cm$^2$.
The effect of clipping is small for the present study.
For the simulations used in this work, the size of the correction is at most 10\,\% of the total radiation energy, and for the majority of the simulations it is smaller than 2\,\%. 
Clipping will be more relevant for higher energies or for observatories at higher altitude.
\par 
In summary, in this work $S_{\textrm{RD}}^{\rho_{\textrm{\Xmax}}}$ is calculated from the electromagnetic shower energy of the full shower by Eq.~\ref{eq:Srd}.
Then, $E_{\textrm{rad}}$ is calculated using Eq.~\ref{eq:Erad}, and the radiation energy $S_{\textrm{RD}}^{\rho_{\theta} \prime}$ is corrected for the atmospheric density depending on the zenith angle by Eq.~\ref{eq:SrdZenUnclipped}.
Finally, the radiation energy is clipped according to the observation level by Eq.~\ref{eq:SrdZen} to gain \Srd used as observable in the present work. 



\end{document}